\documentstyle[twoside,fleqn,psfig,epsf,espcrc2]{article}%
\begin{document}                                                        
\renewcommand{\refname}{\normalsize\bf References}
\title{%
``Quantum phase transitions'' in classical nonequilibrium processes
}
\author{%
       Eldad Bettelheim$^a$,
	 Oded Agam %
\address{Department  of Physics, The Hebrew University Jerusalem, Israel 91904}
\thanks{ This research was supported by THE ISRAEL SCIENCE FOUNDATION founded by
The Israel Academy of Science and Humanities, and by
Grant No.~9800065 from the USA-Israel   
Binational Science Foundation (BSF)\,.},
	\,Nadav M. Shnerb %
\address{Department   of Physics, College of Judea and Samaria, Ariel}%
\thanks{ N. M. S. acknowledges the support of the Valazzi-Pikovsky Fellowship Fund\,.},
}

\begin{abstract}
\hrule
\mbox{}\\[-0.2cm]

\noindent{\bf Abstract}\\
Diffusion limited reaction of the Lotka-Volterra type 
is analyzed taking into account the discrete nature of the 
reactants. In the continuum approximation, the dynamics is dominated by 
an elliptic fixed-point. This fixed-point becomes unstable due to
discretization effects, a scenario similar to quantum phase transitions. As a
result, the long-time asymptotic behavior of the system changes and the
dynamics flows into a limit cycle.
 The results are verified by numerical simulations.
 \\[0.2cm]
\hrule
\end{abstract}
\maketitle
\section{Introduction}
Nonequilibrium systems of diffusing reactants are very common in nature.  
In chemistry almost any chemical reaction is a reaction-diffusion system. 
In physics, the standard examples are annihilation of electrons and holes 
moving in a disordered media, or vortices and 
antivortices in type two superconductors. Examples from other fields include: 
population dynamics in biology, spread of epidemics in health science, and 
group decision dynamics in social science.   
 
It is customary to denote the various types of reactants by capital letters, 
 $A$, $B$, $C$, etc., and the rates of the reactions by Greek 
letters $\mu$, $\lambda$, $\sigma$, etc. Then, for example, a process 
in which $A$ and $B$ annihilate 
each other at rate $\lambda$ is represented symbolically as: 
$ A  + B  \stackrel{\lambda}{\longrightarrow} \emptyset$.   
Similarly, a process where the reaction of $A$ and $B$ produces 
 $C$, at rate $\mu$, is represented by $A  + B   
\stackrel{\mu}{\longrightarrow} C$.  
 
The simplest description of reaction-diffusion dynamics employs the 
densities of the reactants as the basic ingredients of the equations  
of motion. For example, the equation describing the 
binary annihilation reaction  
\begin{equation} 
A + A  \stackrel{\mu}{\longrightarrow} \emptyset \label{AA} 
\end{equation} 
is  
\begin{eqnarray} 
\frac{\partial n_A}{\partial t}= D \nabla^2  n_A - \mu n_A^2, \label{AAE} 
\end{eqnarray} 
where $n_A$ is the density of the reactants, and $D$ is their diffusion 
constant. The first term in the above equation represents the diffusive 
behavior of the particles, while the second term accounts for the 
interaction. We shall call these kind of  
equations ``mean field equations'' for reasons which will be clarified 
later on.  
 
The evolution of many nonequilibrium processes are adequately  
described by mean field equations. One of  
the most remarkable examples is the  Belousov-Zhabotinskii reaction 
\cite{murray} where  
a mixture of few chemical reactants produces a nonequilibrium  
process which is periodic in time.  
 
Facing the success the the mean field theory,
it is natural to ask whether, indeed, it always gives an accurate 
description of reaction-diffusion systems.
In fact, it is known that
the answer for this question is negative. Deviations from the mean 
field theory appear, usually, in systems of low dimensionality.
Returning to the example (\ref{AA}), it can be easily seen 
that the homogeneous solution of equation (\ref{AAE}) behaves 
asymptotically as $n_A \sim 1/t$, 
independent of the dimensionality of the system.  
However, the true asymptotic behavior of (\ref{AA}) is  
$n_A \sim 1/\sqrt{t}$ in $1 d$,  $n_A \sim \ln t/t $ in $2 d$, 
and $n_A \sim 1/t$, for $d>2$ \cite{wilczek}. 

$d=2$ is the critical dimension for 
reaction-diffusion type of nonequilibrium processes. 
The qualitative explanation for this behavior is 
clear: In order to react, the two particles should first diffuse to make 
contact. This diffusion time restricts the rate of the reaction since, 
in low dimensions, diffusion is inefficient  
in mixing the reactants. The Ovchinnikov-Zeldovich segregation  
phenomenon \cite{Ovchinnikov} is the result of spatially inactive regions  
developed in diffusion limited reactions. 
 
The purpose of this work is to present an example for nonequilibrium 
process in which the discretized nature of the reactants has a strong 
impact on the behavior: The mean field equations  
fail to describe the dynamics of the system in the long time
limit. Discretization effects lead to a different asymptotic behavior of the
system in the long time limit. This change is, in a sense, analogous to a 
quantum phase transition.

The prototype example we shall use is one of the simplest models 
in population biology: the Lotka-Volterra  system \cite{lotka,volterra}.
 In the predator-prey 
version of this model, two species, a predator (A) and a prey (B) are 
interacting while all other environmental factors are assumed  intact. 
In the absence of predator, the prey population grows exponentially, 
while in the absence of prey the predator death rate results in an 
exponential decay of their population. 
Binary interaction between the species involves the growth of the predator 
population due to consumption of the prey, thus: 
\begin{eqnarray} 
A   \stackrel{\mu}{\longrightarrow} \emptyset, ~~~B   
\stackrel{\sigma}{\longrightarrow} 2 B, 
~~\mbox{and}~~ A + B \stackrel{\lambda}{\longrightarrow} 2 A,  \label{LVP} 
 \end{eqnarray} 
where $\mu$ is the predator death rate, $\sigma$ is the prey birth rate, 
and $\lambda$ is the probability for a predator to eat a prey at the same 
spatial location. We assume that birth of a new predator follows an 
``eating event" at the same site. The mean field equations 
of this system are the Lotka-Volterra equations: 
\begin{eqnarray} 
\frac{d n_A}{dt} = D\nabla^2 n_A - \mu n_A  + \lambda n_A n_B, \nonumber \\ 
\frac{d n_B}{dt} = D\nabla^2 n_B + \sigma n_B  - \lambda n_A n_B. \label{LVE} 
\end{eqnarray} 

\begin{figure}[ht]
\vspace{-0.2in}
\begin{center}
\leavevmode
\epsfxsize=6.0cm
\epsfbox{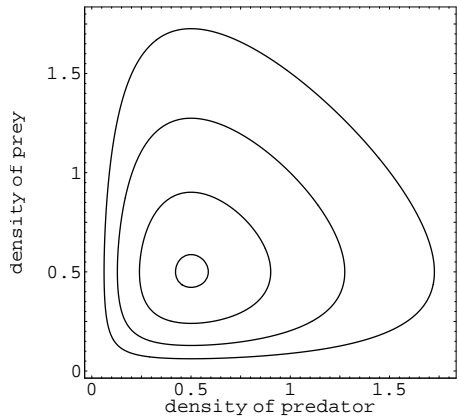} 
\end{center}
\vspace{0cm}
\caption{Examples of  phase space trajectories corresponding 
to the homogeneous solution of  the mean field equations (4).}
\vspace{1cm}
\label{loops}
\end{figure}

\noindent
Here $n_A$ and $n_B$ denote the population densities of the predator and prey,
respectively, and it is assumed that the diffusion constants of both species 
equal to $D$. The generic behavior of this system is periodic in time, 
see Fig. \ref{loops}: When the number prey is large, 
the predator population is growing due to the availability of food, but 
then the prey population decreases. Consequently, also  
the predator population diminishes. When the predator population is already
small, the number of prey, again, begins growing and the cycle repeats. 
  
Here, we will show that the behavior of the quantized version of this model, 
in two dimensions, does not follow the Lotka-Volterra equations (\ref{LVE}). 
Our analysis will proceed in the following way. 
First, we write down the exact  
Master equations of the quantized version of the system. Then, 
we map these equations onto a  
a Schr\"{o}dinger equation in imaginary time, and 
identify the corresponding many-body Hamiltonian. Next, we 
express the propagator of the system  as a field integral and find the 
corresponding action. It will be shown that the mean field equations 
(\ref{LVE})  are the saddle point equations associated with this action.  
The effective action of the system will be, then, constructed following the
traditional procedure of renormalization. Namely, the fields will 
be separated  into  ``fast" and ``slow" components, and 
the fast components will be integrated out. 
Finally, we analyze the saddle point 
equation of the effective action and characterize the long time asymptotic
behavior of the system. 
Our results will be verified by numerical simulations.  

To avoid cumbersome algebraic manipulations, our discussion will 
be switching between two processes: the binary annihilation (\ref{AA}), and 
the Lotka-Volterra reaction (\ref{LVP}).  
The first example will be used as a simple illustration of the 
derivation, while the results for
the Lotka-Volterra reaction will be usually stated without an explicit 
derivation.
 
\section{The Master Equations} 
 
The Master equations of a nonequilibrium process are equations for 
the probabilities of the various states of the system. Consider 
the binary annihilation process (\ref{AA}) in zero dimension,  i.e. 
when the system consists of a single site. Then, a state of the system
is defined by the number of reactants, $n$, and we denote 
the probability to find the system in this state by $P_n$. 
The Master equations relate the change of probability in time to the
rate of flow into and out of the state:
\begin{eqnarray} 
\frac{d P_n}{dt} =  -\frac{\mu}{2} \left[ n(n\!-\!1) P_{n}\!-\!
(n\!+\!2)(n\!+\!1) P_{n+2}\right]. \label{AAM}
\end{eqnarray} 
The first term on the right hand side represents 
the flow out of the state with $n$ reactants. It
comes from pair annihilation, and therefore proportional  
to the number of pairs, $n(n-1)/2$.
The second term accounts for the flow into the state
which is due to pair annihilation in the state with $n+2$ reactants.

The same logic can be used  in order to construct the Master 
equations for the Lotka-Volterra process (\ref{LVP}). 
Considering again the zero dimensional case, they take the form
\begin{eqnarray} 
\frac{ d P_{m,n}}{d t} 
 &=&  - (\mu m \!+\! \sigma n \!+\! \lambda m n) P_{m,n} \label{LVME} \\ 
&+& \mu (m\!+\!1) P_{m+1,n}\! +\!   
\sigma (n\!-\!1) P_{m, n-1} \nonumber \\
&+& 
\lambda (m\!-\!1)(n\!+\!1) P_{m-1,n+1}, \nonumber
\end{eqnarray} 
where $P_{m,n}$ denotes the probability 
to find  the system in a state with $m$ predator and $n$ prey.

The generalization of the above equations to the non-zero dimensional 
case is straightforward. A state of the system is now defined by two vectors
of integer numbers: ${\bf n}  = (n_1,n_2,\cdots)$,
and  ${\bf m } = (m_1,m_2,\cdots)$. The components of these vectors represent 
the occupation numbers of  the prey and the predator at the various
 sites of the system, 
and $P_{{\bf m},{\bf n}}$ is  the joint probability of the 
occupation configurations.
The Master equations, in this case, contain an additional 
hopping term between the sites. 
This term will be added to our theory later on. 

\section{Mapping the Master equations onto a Schr\"{o}dinger equation}

We turn now to map the Master equations 
onto the Schr\"{o}dinger equation in imaginary time, and to identify the  
corresponding many-body Hamiltonian. 
Beginning with the example of binary annihilation process  
in zero dimensions, following Refs.~\cite{peliti,Cardy}, 
we define the wave function    
\begin{eqnarray}
| \psi \rangle = \sum_{n = 0}^\infty P_n | n \rangle,
\end{eqnarray}
where $| n \rangle$ denotes a state with $n$ reactants in the system. 
Taking the derivative 
of $| \psi \rangle$ with respect to time and substituting 
(\ref{AAM}) we obtain
\begin{eqnarray}
\frac{d}{d t} | \psi \rangle = \sum_n \frac{d P_n}{d t} | n \rangle =
~~~~~~~~~~~~~~~~~~~~~~~~~~~~\label{AAI}\\
 -\frac{\mu}{2} \sum_n \left[   n(n\!-\!1) P_n\!-\!(n\!+\!2)(n\!+\!1) P_{n+2}
\right] |n \rangle. \nonumber
\end{eqnarray}
Let us now introduce the
creation, $\hat{a}^\dagger$, and annihilation, $\hat{a}$, operators 
which satisfy the Bose commutation 
relation $[\hat{a}, \hat{a}^\dagger]=1$, and
\begin{eqnarray}
\hat{a}^\dagger |n \rangle = |n +1 \rangle, ~~~~\mbox{while} ~~~
\hat{a} |n \rangle = n|n -1 \rangle.
\end{eqnarray}
It is easy to see that $\hat{a}^2 |\psi \rangle = 
\sum_n P_{n+2} (n+2)(n+1) |n \rangle$, and
$(\hat{a}^\dagger)^2 \hat{a}^2 |\psi \rangle = 
\sum_n P_{n} (n-1)n |n \rangle$. Substituting these results in equation
(\ref{AAI}), one can write it in the from of a Schr\"{o}dinger 
equation in imaginary time
\begin{eqnarray}
\frac{d}{d t} | \psi \rangle = - H | \psi \rangle, \label{SE}
\end{eqnarray}
where the Hamiltonian is given by 
\begin{eqnarray}
H = \frac{\mu}{2}(\hat{a}^\dagger\hat{a}^\dagger-1)\hat{a}\hat{a}. \label{HA}
\end{eqnarray}

Turning to the Lotka-Volterra reaction in zero dimension, we denote by
$|m,n\rangle$ the state with $m$ predator and $n$ prey. The corresponding
wave function is $| \psi \rangle = \sum_{n,m} P_{m,n} |m, n \rangle$,
and the equations of motion take the same form as (\ref{SE}), but with 
the Hamiltonian:
\begin{eqnarray}
H= \mu (\hat{a}^\dagger\!-\!1) \hat{a} \!+\! \sigma ( 1\!-\! 
\hat{b})\hat{b}^\dagger \hat{b} 
\!+\! \lambda \hat{a}^\dagger (\hat{b}^\dagger\!-\!\hat{a}^\dagger)  
\hat{a} \hat{b}. \label{HLV}
\end{eqnarray}
Here $\hat{a}^\dagger$ and $\hat{a}$ are the creation and annihilation 
operators of predator, while  $\hat{b}^\dagger$ and $\hat{b}$ are the 
creation and annihilation operators of prey. 

The generalization of equation (\ref{SE}) to the nonzero dimensional case 
is obtained by defining the creation and annihilation operators at each site
of the system (i.e.~$ \hat{a}^\dagger \to \hat{a}^\dagger_i, 
\hat{a} \to \hat{a}_i, \hat{b}^\dagger \to \hat{b}^\dagger_i , 
\hat{b} \to \hat{b}_i, \mbox{with}~ 
[\hat{a}_j, \hat{a}^\dagger_i] = \delta_{ij}$, etc.),  and adding a hopping 
term to the Hamiltonian. The wave function, in this case, is
\begin{eqnarray}
 | \psi \rangle = \sum_{{\bf n},{\bf m}} P_{{\bf m},{\bf n}}
\prod_i (\hat{a}^\dagger_i)^{m_i} (\hat{b}^\dagger_i)^{n_i} |0 \rangle, 
\nonumber
\end{eqnarray}
where $m_i$ and $n_i$ are the components of ${\bf m}$ and ${\bf n}$, and
$|0 \rangle$ denotes the vacuum state with
no reactants in the system.

\section{The formal solution of the Schr\"{o}dinger equation, and 
expectation values}

The formal solution of the Schr\"{o}dinger equation (\ref{SE}) is
\begin{eqnarray}
| \psi (t) \rangle = U(t) | \psi (0) \rangle, \nonumber
\end{eqnarray}
where  $\psi (0)$ is the initial wave function, and $U(t)$ is the propagator
of the system for time $t$, i.e.
\begin{eqnarray}
U(t) = {\cal T} \exp \left\{ - \int_0^t d t' H(t') \right\}, \nonumber
\end{eqnarray}
${\cal T}$ being the time ordering operator. 
This solution of $\psi$, as function of the time, 
fully characterize the behavior of the system. 
Notice, however, that the interpretation of the 
wave function differs from that of quantum mechanics. 
Here it represents probability and not probability 
amplitude. In particular, the expectation value of an operator, $\hat{Q}$, 
in a state defined by $\psi$ is given by the matrix 
element \cite{peliti,Cardy} 
\begin{eqnarray}
\langle \hat{Q} \rangle = \langle {\cal P} | \hat{Q} | \psi \rangle \label{AQ}
\end{eqnarray}
where 
\begin{eqnarray}
|{\cal P} \rangle = \prod_{i} e^{ \hat{a}^\dagger_i + \hat{b}^\dagger_i}
 | 0 \rangle, \nonumber
\end{eqnarray}
is an eigenstate of the annihilation operators, i.e. 
\begin{eqnarray}
\hat{a}_j |{\cal P} \rangle = \hat{b}_j |{\cal P}=|{\cal P}\rangle
~~~\mbox{for any}~~j.
\label{CS}
\end{eqnarray}

In understanding the structure of this nonequilibrium theory, it
is instructive to consider specific examples of expectation values.
Consider, first, the expectation value of the identity operator $\hat{Q}=1$. 
Using the identity 
\begin{eqnarray}
e^{ \hat{a}} f(\hat{a}^\dagger, \hat{a})= f(\hat{a}^\dagger+ 1, \hat{a}) 
e^{ \hat{a}}, \label{OI}
\end{eqnarray}
where $f(\hat{a}^\dagger, \hat{a})$ 
is a general function of the creation and annihilation operators,
one can easily see  that $\langle {\cal P}| \psi \rangle= 
\sum_{{\bf n},{\bf m}} P_{{\bf m},{\bf n}} $.
Since the sum of probabilities over all possible occupation configurations
equals unity, we obtain $\langle \hat{Q} \rangle =1$. 
Thus the normalization of the wave function reads 
$\langle P| \psi \rangle =1$. 

The conservation of probability implies that 
$\langle P| \psi \rangle =1$ holds for any time $t$, therefore
\begin{eqnarray}
\frac{d}{dt} 
 \langle {\cal P} | U(t) |\psi \rangle =0. \nonumber
\end{eqnarray}
This equation is satisfied only if 
$\langle {\cal P} | H =0$. Thus,
a legitimate Hamiltonian of our theory must vanish when setting all 
the creation operators to one. For example, the Lotka--Volterra Hamiltonian
should satisfy
\begin{eqnarray}
 H(\{ \hat{a}_i^\dagger=1, \hat{b}_i^\dagger=1,\hat{a}_i, \hat{b}_i\}) =0.
\nonumber 
\end{eqnarray}
It is easy to verify that (\ref{HA}) and (\ref{HLV}), indeed, 
satisfy this condition. 
 
As a second example, let us calculate the mean 
number of predator at site $i$, 
$\bar{n}_i= \langle \hat{a}^\dagger_i \hat{a}_i \rangle$. Notice 
that, unlike quantum mechanics where the expectation value of
an annihilation operator vanish, here (\ref{CS}) implies that  
$\bar{n}_i= \langle \hat{a}_i \rangle$. 
Substituting $\hat{Q}= \hat{a}_i$ in (\ref{AQ}) and using (\ref{OI}) we obtain 
$\bar{n}_i=\sum_{{\bf n},{\bf m}} n_i P_{{\bf n},{\bf m}}$ where $n_i$ 
is the number of prey at site $i$.

\section{The 
field theoretic formalism}

In order to construct the propagator, $U(t)$, it is convenient to
employ the path integral formalism. To begin with, let us consider
the propagator of binary annihilation in the zero 
dimension.  We define a coherent state of the system as 
\begin{eqnarray}
| a \rangle = e^{- \frac{|a|^2}{2} } e^{a \hat{a}^\dagger} | 0 \rangle,
\nonumber
\end{eqnarray}
where $a$ is a complex number. The matrix element of a normal
ordered operator $f(\hat{a}^\dagger, \hat{a})$, where all creation  
operators stand left to  annihilation operators, is given by
\begin{eqnarray}
\langle a |f(\hat{a}^\dagger, \hat{a})| a' \rangle =
f(a^*,a') e^{-\frac{1}{2}( | a^2+ |a'|^2 - 2 a^* a')}, \label{afa}
\end{eqnarray}
In particular, the  inner product of coherent states is 
$\langle a | a' \rangle = \exp 
\{-\frac{1}{2}( | a|^2+ |a'|^2 - 2 a^* a' )\}$, and the normalization
condition $\langle a | a \rangle =1$ is satisfied. 
The resolution of identity associated with coherent states is
\begin{eqnarray}
\int\frac{ d^2 a }{\pi}~~ | a \rangle \langle a |, \label{ROI}
\end{eqnarray}
where $d^2 a= d \Re a ~d \Im a$.

Consider the expectation value of a general operator 
$\hat{Q}=Q(\hat{a}^\dagger,\hat{a})$ at time $t$, 
\begin{eqnarray}
\langle \hat{Q}(t)\rangle = \langle P | Q(\hat{a}^\dagger,a) 
e^{-H(\hat{a}^\dagger, a) t }|\psi \rangle, \nonumber
\end{eqnarray}
where operators are assumed to be normal ordered.
For simplicity we choose an initial state, 
$|\psi \rangle = \sum_n P_n (\hat{a}^\dagger )^n | 0 \rangle$,
with $P_n= e^{-1}/n!$, thus 
$|\psi \rangle = e^{\hat{a}^\dagger-1}| 0 \rangle$. Since we are 
interested in properties which are independent of the precise form of
the initial condition, this particular choice does not have an
important effect. 

>From (\ref{OI}) and general properties of the propagator 
it follows that 
\begin{eqnarray} 
\langle \hat{Q}(t)\rangle = 
\langle 0 | Q(1,\hat{a}) \times~~~~~~~~~~~~~~~~~~~~~~~~~~~~~~~~~ \nonumber \\ 
  ~~~~~~~~~\left[e^{-H(\hat{a}^\dagger+1, a) \Delta t }
\cdots e^{-H(\hat{a}^\dagger+1, a) \Delta t } \right]
e^{\hat{a}^\dagger}|0 \rangle, \nonumber
\end{eqnarray}
where the square brackets contain a product of $N$ infinitesimal 
propagators for times $\Delta t = t/N$.
Now, we insert $N+1$ identity  operators (\ref{ROI}) between the various 
terms of the above product. Using  
(\ref{afa}) and taking the continuum limit,
$N \to \infty$, we obtain the expectation value $\langle Q(t)\rangle$
in the form of a path integral
\begin{eqnarray} 
\langle Q(t)\rangle =  \int{\cal D}[ a^*, a ] e^{-F_0}~ Q(1, a(t))  
e^{-|a(0)|^2+ a^*(0)} \nonumber
\end{eqnarray}
where $F_0$ is the action of the system
\begin{eqnarray}
F_0= \int_0^t d\tau \left\{ a^*(\tau) \partial_\tau a(\tau) 
+ H[a^*(\tau)+1,a(\tau)]\right\}, \nonumber
\end{eqnarray}
and ${\cal D}[ a^*, a ] = \prod_\tau [d^2 a(\tau)/\pi]$ is
the measure of the integral.

Having the path integral expression for the propagator,  
the generalization to the finite dimensional case
is straightforward. It merely amount for the addition of a 
diffusive term in the action. Thus the action of the diffusion-reaction
process (\ref{AA}) is
\begin{eqnarray}
F= \!\int \!d{\bf r}d\tau~ \left\{a^* (\partial_\tau \! -\! D\nabla^2)
a\!+\! \frac{\mu}{2} [a^{*2} a^2\!-\!2 a^*a^2]\right\}, \nonumber
\end{eqnarray}
where, now, $a^*(r,\tau)$ and $a(r,\tau)$ are functions of the time $\tau$
as well as the space coordinates ${\bf r}$. 
Henceforth we omit the explicit time and space 
dependence of the fields.

The action associated with the Lotka-Volterra process (\ref{LVP}) 
can be derived in the same way, and the result is:
\begin{eqnarray}
F\!=\!\! \int   a^*( \!\partial_\tau \!-\! D \nabla^2 ) 
 a \!+\!b^* (\partial_\tau \!-\! D \nabla^2 )  b \!+\!
 \tilde{H}, \label{action}
\end{eqnarray}
where
\begin{eqnarray}
\tilde{H} = \mu a^* a -\sigma (1+b^*) b^*b+ \lambda (a^*+1)(b^*-a^*)a b.
\nonumber
\end{eqnarray}

Finally we remark that, unlike quantum mechanics of many-body bosons
where fields are periodic in imaginary time, here there are no 
such boundary conditions.

\section{The mean field equations}

The mean field equations of the field theory defined by the action
(\ref{action}) are the saddle point equations of the bare action, 
\begin{eqnarray}
\delta F =0, \nonumber
\end{eqnarray}
where the functional derivative is with respect to all the fields,
$a,~ a^*, b$ and $b^*$. Seeking for a solution with non vanishing
densities $\bar{n}_A=\langle \hat{a} \rangle$,  
$\bar{n}_b=\langle \hat{b} \rangle$,
the saddle fields (which we denote by bar) are  $\bar{a}^*=\bar{b}^*=0$,
and solutions of the equations
\begin{eqnarray} 
\frac{d \bar{a}}{d\tau} = D\nabla^2 \bar{a} - \mu \bar{a}  +
 \lambda \bar{a}\bar{ b}, \nonumber \\ 
\frac{d \bar{b}}{d\tau} = D\nabla^2\bar{ b} + \sigma \bar{b} 
 - \lambda \bar{a} \bar{b}. \nonumber 
\end{eqnarray} 
Thus the mean filed equations of the quantized Lotka-Volterra 
system are precisely the Lotka-Volterra equations (\ref{LVE}). 

The steady state solutions of (\ref{LVE}), $dn_A/dt=dn_B/dt=0$,
admit only homogeneous densities in space \cite{okubo}.
They are associated with the fixed-points of the equations: 
One is the unstable hyperbolic fixed-point, $n_A \!=\!n_B \!= \!0$, 
corresponding to the case  with no reactants in the system.
The second, $n_A\!=\!0,~ n_B\!=\!\infty$, represents the situation 
where the number of prey grows indefinitely in the absence of predator. 
The third fixed-point, $\bar n_A \!=\! \sigma / \lambda$,   
$\bar n_B \!=\! \mu/ \lambda$, is an elliptic fixed-point
corresponding to a balanced ecological state with
fixed populations. Linearizing equations (\ref{LVE}) around 
the latter fixed-point we obtain:
\begin{eqnarray}
\frac{d}{dt} \left( \begin{array}{c}
\Delta n_A \\  \Delta n_B
\end{array} \right) \simeq  -  M_0 \left( \begin{array}{c}
\Delta n_A \\  \Delta n_B
\end{array} \right),
\end{eqnarray}
where
\begin{eqnarray}
M_0 = 
 \left( 
\begin{array}{cc}
0  & - \sigma   \\
 \mu & 0   \\
\end{array}
\right), \label{M0}
\end{eqnarray}
$\Delta n_A= n_A- \bar{n}_A$, and $\Delta n_B= n_B- \bar{n}_B$.

The matrix $M_0$, which we shall call the bare mass matrix, 
is the generator of time evolution
of homogeneous densities in the vicinity of the fixed-point. Its eigenvalues
determine the stability properties of the fixed-point.  If the real
parts of the eigenvalues are positive, the fixed-point is stable,
while if negative, it is unstable. In our case, 
the eigenvalues of $M_0$ are purely imaginary, 
$\epsilon_\pm=\pm i \sqrt{\mu \sigma}$.
It implies that close enough to the fixed-point the population 
densities oscillate in time. Moreover, a nonlinear stability analysis
of the Lotka-Volterra equations shows that
the system exhibits periodic evolution over the whole phase space.
The reason is the existence of a conserved quantity\cite{murray} 
(for homogeneous densities),
\begin{equation}   
K = n_A + n_B - {\mu \over \lambda}  \ln (n_B) - {\sigma \over \lambda}
\ln (n_A),
\end{equation}
which confines the phase space trajectories  
to move along concentric closed loops, as shown in Fig.~\ref{loops}. 

\begin{figure}
\vspace{-0.4in}
\begin{center}
\leavevmode
\epsfxsize=7.7cm
\epsfbox{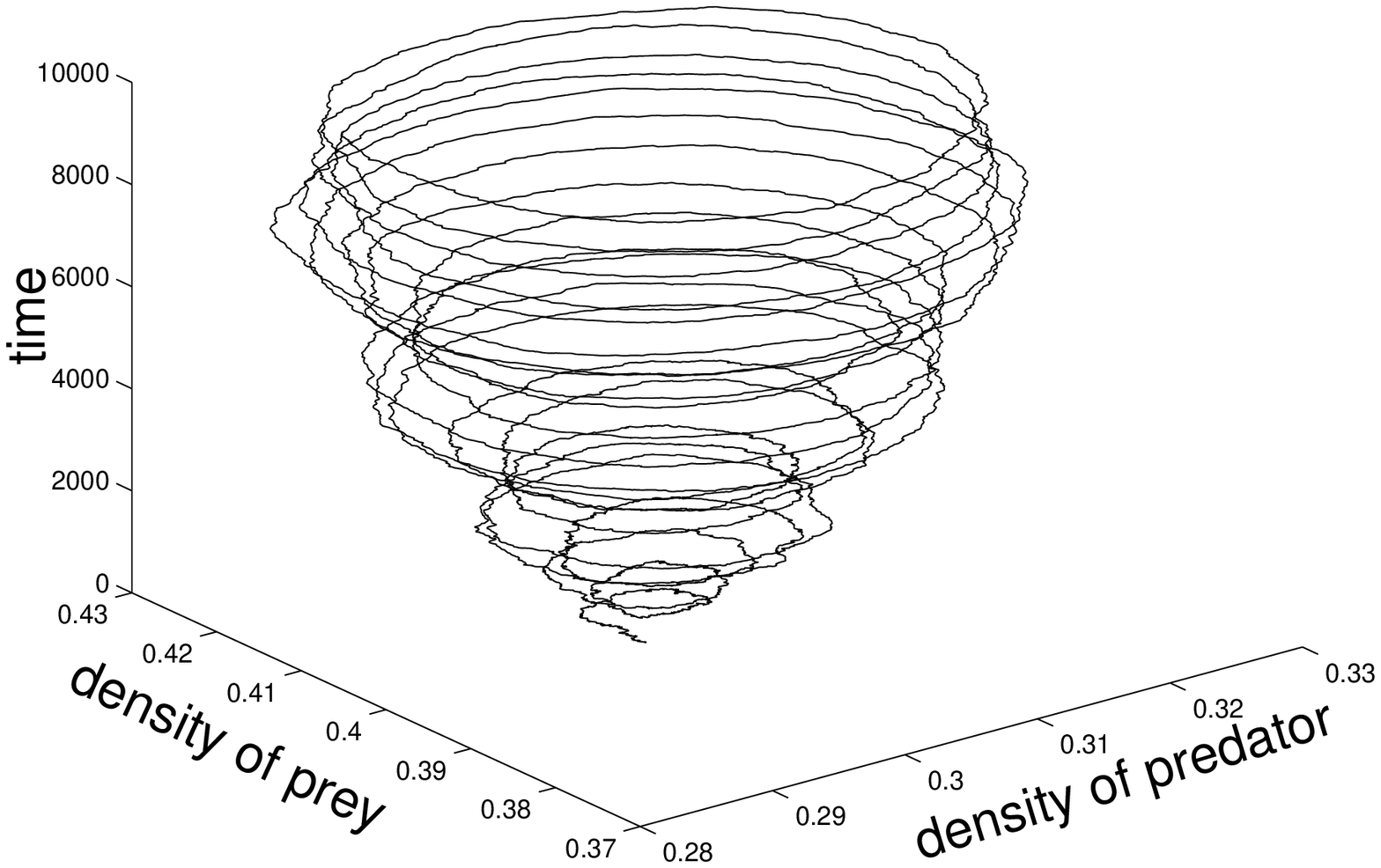} 
\end{center}
\vspace{0in}
\begin{center}
\leavevmode
\epsfxsize=7.7cm
\epsfbox{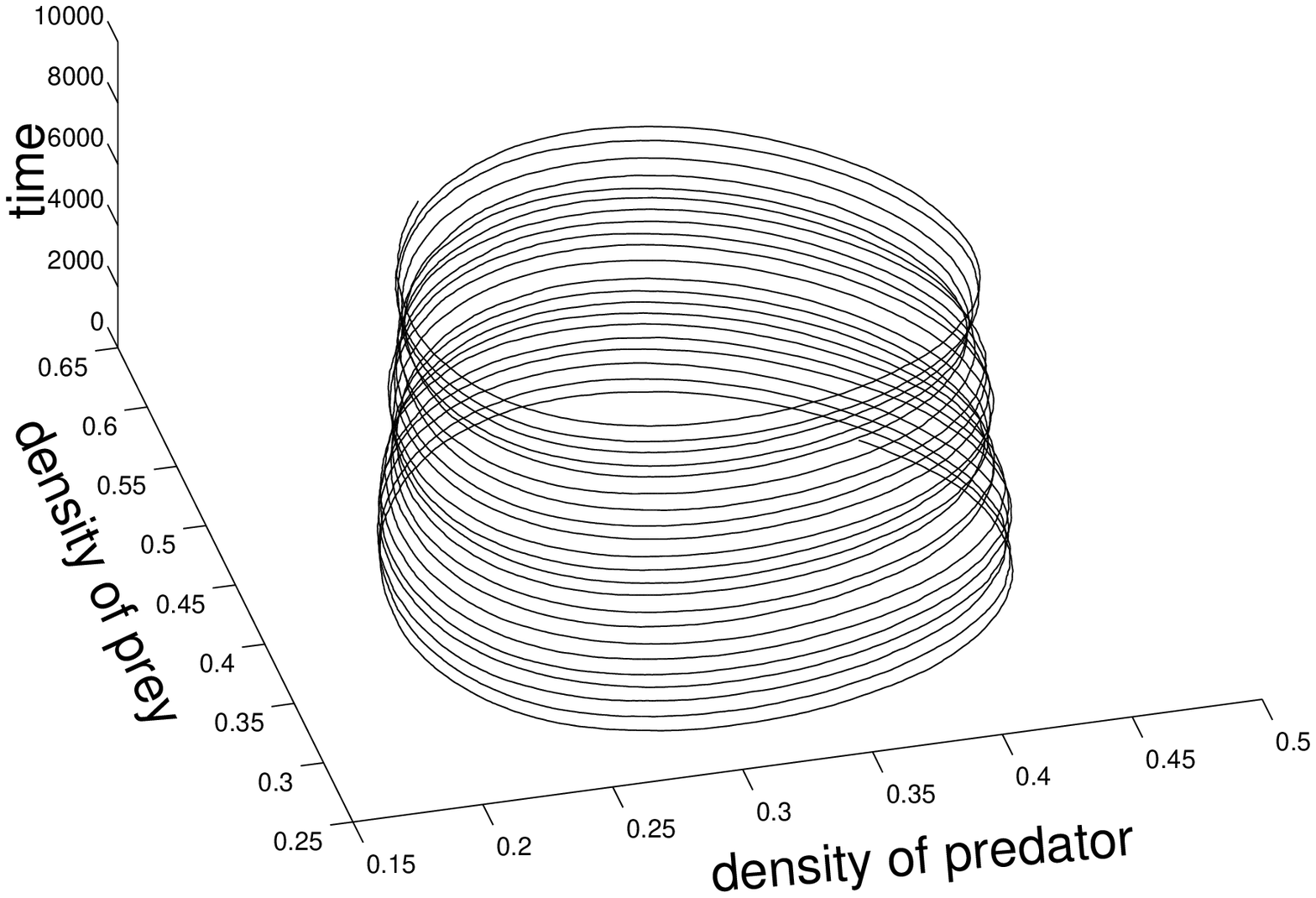} 
\end{center}
\vspace{-0.3in}
 \caption{Plot of prey and predator densities as a function of time for $\lambda=0.02, \sigma=0.006, \mu=0.008$. Upper panel
 shows simulations with initial conditions near the fixed point. Here the system flows away from the fixed point. The lower panel shows the flow when the initial conditions are set to be far from the fixed point. The converging behavior of the trajectories signals that the dynamics is attracted to a limit cycle. }
\vspace{-0.3cm}
\label{simulations}
\end{figure}

Notice that elliptic fixed-points are unstable 
with respect to  small perturbations. Any small 
perturbation might shift the eigenvalues $\epsilon_\pm$ off 
the imaginary axis. Consequently the elliptic fixed-point will
become either
a stable or an unstable focus. In what follows it will be shown that
discretization effects, which lie beyond the mean field description,  
indeed, lead to such a scenario.

\section{Renormalization procedure} 

An effective action of a field theory is obtained by 
integrating out the ``fast'' degrees of freedom. The procedure is usual:  
First, we separate the fields into ``fast'' and ``slow'' components. Say
for the annihilation process (\ref{AA}),  
$a=a_{f}+ a_{s}$, where $a_f$ and $a_s$ denote Fourier components 
of the field which oscillate rapidly or slowly in space, respectively. 
Next, we expand the action $F[a, a^*]$ up to 
second order in the fast fields:
$F\simeq 
F[a_s, a^*_{s}] + F_2[a_{s}, a^*_{s}; a_{f},a^*_{f}]$,
where  $F_2$ is quadratic in the fast fields 
$a_{f}$ and $a^*_{f}$.
Then the effective action, $F_{eff}[a_{s}, a^*_{s}]$, is 
obtained by integrating out the fast fields,  
\begin{eqnarray}
e^{-F_{eff}}
= \int {\cal D} [a_{f}, a^*_{f}] 
e^{-F[a_{s}, a^*_{s}] - F_2[a_{s}, a^*_{s}, a_{f},a^*_{f}] }.\nonumber
\end{eqnarray}

For the binary annihilation reaction, the effective action
has been calculated by Cardy and Tauber \cite{Cardy}. From 
this calculation they
prove that, for $d>2$,
the mean field equations hold, and the density of reactants decreases as
$n(t)  \sim 1/t$. However, when $d<2$, 
inherent  spatial fluctuations result in a 
decrease of the asymptotic annihilation rate,  
$n(t)  \sim 1/t^{d/2}$ \cite{wilczek}.

Notice, however, that in the binary  annihilation example
the fixed-point $n_A( t \to \infty)=0$ does not change due 
to discretization effects. Only the 
asymptotic approach to this point alters.
As we shall see, in the Lotka-Volterra system the renormalization 
procedure results in a more
dramatic effect. The nature of the fix-point itself changes.
To put it differently, if one associates the long time asymptotic behavior
of the system with the ground state of the field theory defined by
the action (\ref{action}), then quantization leads to a new ground state
which differ from the ``classical'' ground state described by the 
Lotka-Volterra
equations (\ref{LVE}). 

In what follows we consider only the physical situation of the
critical dimension, $d=2$, and defer the technical details of the derivation
of the effective action of the Lotka-Volterra system to the Appendix.

Given the fixed-point of the Lotka-Volterra equations (\ref{LVE})
at finite densities,  $(\bar{n}_A, \bar{n}_B)= (\frac{\sigma}{\lambda}, 
\frac{\mu}{\lambda})$, it is convenient to change variables to
fields which represent fluctuations around this
fixed-point, namely, $a \to a + \bar{n}_A$, and 
$b \to b + \bar{n}_B$. The bare Green 
function, $G^0$, associated with the quadratic part of the resulting 
action is, 
\begin{equation}
G^0 = \left[ 
(\partial_t-D \nabla^2)\tau_0 + M_0 \right]^{-1}, \label{G0}
\end{equation}
where $\tau_0$ is the identity matrix, and $M_0$
is the bare mass matrix (\ref{M0}).
Dyson's equation for the exact Green function, $G$, is, 
\begin{equation}
G= G^0 + G^0 \Sigma G,
\end{equation}
where $\Sigma$ is the self energy. The correction for
the mass matrix is the zero Fourier components 
of the self energy, $\delta M=\Sigma(k=0,\omega =0)$, thus $M=M_0+\delta
M$. The stability of the fixed point is determined by the eigenvalues of the
renormalized mass matrix, which are 
\begin{equation}
\epsilon_\pm=\frac{1}{2}\left(\mbox{Tr}(M)\pm\sqrt{\mbox{Tr}(M)^2-
4\mbox{Det}(M)}\right).
\end{equation}
Renormalization, in the first approximation, only shifts $\mbox{Tr}(M)$ from its
zero mean field value. Thus the eigenvalues of the mass matrix are
approximately $\epsilon_\pm\approx\mbox{Tr}(M)/2\pm
i\sqrt{\mu\sigma}$, where 
\begin{eqnarray}
\mbox{Tr}(M)=\left.\frac{\lambda}{16\pi
D}\right(\left(6+\log(16)\right)\left(\mu-\sigma\right)+\\
\left.\left(\pi+2\right)\frac{\sigma^\frac{3}{2}}{\mu^{\frac{1}{2}}}+\left(\pi-2\right)\frac{\mu^\frac{3}{2}}{\sigma^{\frac{1}{2}}}+\left(6\pi+8\right)\sqrt{\mu\sigma}\right)\nonumber
.
\label{newtrace}
\end{eqnarray}
In deriving this result (see Appendix) we have assumed $\frac{\lambda}{D}\ll 1$, 
$\log\left(\frac{D\Lambda^2}{\sqrt{\mu\sigma}}\right)\gg 1$,  
$\frac{\lambda}{D}\log\left(\frac{D\Lambda^2}{\sqrt{\mu\sigma}}\right)\ll 1$,
and low densities of the reactants. $\Lambda$ is the upper momentum cutoff. 
>From this result it follows that the fixed point becomes an unstable
focus. This is verified by numerical simulations shown in the upper panel of
Fig.~\ref{simulations}.

It is natural inquire about the nature of the new ground state 
of the system. What is
the global behavior of trajectories in 
phase space when the fixed-point is an unstable focus?
There are two sensible scenarios: (a) One of the other 
mean field fixed-points,  $n_A = n_B =0$ or 
$n_A = 0, n_B = \infty$, becomes stable due discretization,
and all the trajectories converge to this point. 
(b)  Some other type of attractive manifold, 
such as a limit cycle, if formed.
In the lower panel of Fig.~\ref{simulations} we present numerical results of
simulations where we set the  initial conditions to be far from the fixed point. The
converging behavior of the trajectories indicates that the second scenario
takes place, and the long time asymptotics of the system is that of a limit cycle. 

\section{Summary}

We have shown that the discreteness of the reactants in Lotka-Volterra 
systems, in two dimensions, results in a behavior (Figs.~\ref{simulations})
which differs from that of
the Lotka-Volterra equations (Fig.~\ref{loops}). Our analytical analysis and numerical simulations (Fig.~\ref{simulations}) 
indicate that the instability of the Lotka-Volterra equations signals 
the formation of a new ground state where an attractive manifold
similar to a limit cycle is formed. Further studies on this subject 
will be focused on the nature of the unstable phase, and the behavior 
of the system in one dimension.   

\section{Appendix}
 
In this appendix we provide some details of the summation of the one loop diagrams for the Lotka Volterra System.  Our discussion will be limited to properties of the system near the fixed-point. It is therefore convenient to change variables to fields which describe fluctuations around this fixed-point. Thus changing variables as $a \to a + \bar{n}_A$,  $b \to b + \bar{n}_B$,  the action (\ref{action}) takes the form
\begin{eqnarray}
F= \int  \!d{\bf r}   dt  ~ a^* (\partial_t \!-\! D \nabla^2 )  a \!+\!
b^* (\partial_t \!-\! D \nabla^2 )  b \!+\!   \bar{H}, \nonumber
\end{eqnarray}
where 
\begin{eqnarray}
\bar{H} &=& \mu  b^* a \!-\! \sigma a^* b \!+\! \sigma (a^*\!-\!b^*) b^* b 
\!-\! \sigma  a^* a^* b  \nonumber \\
 &+& \mu (b^*\!-\!a^*)a^* a + \lambda (b^*\!-\!a^*) b a  +
\lambda a^* a b^* b \nonumber \\ 
&-&\lambda a^* a^* a b  
\! +\! {\mu \sigma \over \lambda} (b^* a^* \!-\! a^* a^* \!-\! b^* b^* ). 
\nonumber
\end{eqnarray}

To construct the quadratic part of the action
in the fast variables, $F_2$, 
it is convenient to represent the fields in terms of their
Fourier components, e.g. $a({\bf k},\omega)= \int dt d{\bf r}
e^{i {\bf k \cdot r}+ i \omega t} a({\bf r},t)$. Now, we  
define  a vector of fast variables 
$\Psi_f= (a_f,b_f,a^*_f, b^*_f)^T$, where the fast fields contain terms
only with high values of the momentum ${\bf k}$. Thus: 
\begin{eqnarray}
F_2=  \frac{1}{2} \int \frac{d{\bf k} d \omega}{(2 \pi)^3} ~\Psi^\dagger_f
 (Z + G_0^{-1}) \Psi_f, \nonumber
\end{eqnarray}
where elements of the matrix $Z$ are functions of the slow fields
(henceforth we drop the subscript $s$ of these fields), and:
\begin{equation}
G_0^{-1}=\left(\matrix{ D\,q^2 - i\omega & -\sigma  \cr \mu  & D\,q^2 - i\omega \cr  }\right),
\end{equation}
where the bilinear reaction terms have been incorporated into the propagator. 

The integration over the fast variables yields 
$(\mbox{Det}Z)^{-1/2}$ which can be written as 
$ \exp \{- \frac{1}{2}\mbox{Tr} \ln Z \}$.  Thus the
effective action takes the form:
\begin{eqnarray}
F_{eff} \!=\! F \!+\!\frac{1}{2} \mbox{Tr} \ln G_0^{-1}  \!+\! 
\frac{1}{2} \mbox{Tr} \ln[ 1\!+\! 
G_0 Z]
\label{feff}
\end{eqnarray}

The effective mean field equations are obtained from the effective action by
differentiating with respect to the fields $a^*$ and $b^*$. This
differentiation yields the new equations of motion ($\delta F_{eff}=0$):
\begin{eqnarray}
\left.\frac{\delta F}{\delta
a^*}+\frac{1}{2}\mbox{Tr}\frac{1}{1+G_0 Z}G_0\frac{\delta Z}{\delta a^*}\right|_{a^*=b^*=0}=0,\nonumber\\
\left.\frac{\delta F}{\delta
b^*}+\frac{1}{2}\mbox{Tr}\frac{1}{1+G_0 Z}G_0\frac{\delta Z}{\delta b^*}\right|_{a^*=b^*=0}=0.
\label{corrections}
\end{eqnarray}
Here the trace should be understood as  a summation both over the indices of
the matrices and an over $k$ and $\omega$. The latter summation is
logarithmically divergent, signaling that the long time asymptotics of the system is
different from the mean field equations. In the leading approximation we cut
off this divergence at the point where the perturbation expansion breaks down,
namely $k=\sqrt{\sqrt{\mu\sigma}/D}$. In this approximation the location of the
new fixed-point is shifted to:
\begin{eqnarray}
\bar{n}_a\longrightarrow \frac{\sigma}{\lambda}+\frac{\sigma \,\log (\frac{D\,{\Lambda }^2}{{\sqrt{2\sigma\mu}}})}{2\,D\,\pi },\nonumber\\
\bar{n}_b\longrightarrow \frac{\mu}{\lambda}+\frac{\mu \,\log (\frac{D\,{\Lambda
}^2}{{\sqrt{2\sigma\mu}}})}{2\,D\,\pi },
\label{newfixed}
\end{eqnarray}
where $\Lambda$ is the upper momentum cutoff. Here we assume $\frac{\lambda}{D}\ll 1$, 
$\log\left(\frac{D\Lambda^2}{\sqrt{\mu\sigma}}\right)\gg 1$,  
$\frac{\lambda}{D}\log\left(\frac{D\Lambda^2}{\sqrt{\mu\sigma}}\right)\ll 1$,
and low densities of the reactants, namely $\sigma/\lambda\Lambda^2,\mu/\lambda\Lambda^2<1$.

Next, in order to calculate the mass matrix, we expand the equations of motion
(\ref{corrections}) to second order in $a$ and $b$ and to first order in
$\lambda/D$. The resulting equations are linearized around the new fixed point
(\ref{newfixed}), and the corresponding mass matrix is calculated. The result
for the trace of the mass matrix is given in (\ref{newtrace}).

\end{document}